\documentclass[%
 reprint,
showpacs,preprintnumbers,
 amsmath,amssymb,
prb,
superscriptaddress
]{revtex4-2}

\usepackage{graphicx}
\usepackage{dcolumn}
\usepackage{bm}
\usepackage{physics}
\usepackage{comment}
\usepackage{algpseudocode}
\graphicspath{{./figures/}}

\usepackage{color}

\begin{document}


\title{Maxwell-Vlasov-Uehling-Uhlenbeck (VUU) Simulation for Coupled Laser-Electron Dynamics in a Metal Irradiated by Ultrashort Intense Laser Pulses}

\author{Mizuki Tani}
\email{tani.mizuki@qst.go.jp}
\affiliation{%
 Kansai Institute for Photon Science, National Institutes for Quantum Science and Technology (QST), 8-1-7 Umemidai, Kizugawa, Kyoto 619-0215, Japan
}%
\affiliation{%
 Photon Science Center, Graduate School of Engineering, The University of Tokyo, 7-3-1 Hongo, Bunkyo-ku, Tokyo 113-8656, Japan
}
\affiliation{%
 Department of Nuclear Engineering and Management, Graduate School of Engineering, The University of Tokyo,7-3-1 Hongo, Bunkyo-ku, Tokyo 113-8656, Japan
}

\author{Tomohito Otobe}%
\email[corresponding author:]{otobe.tomohito@qst.go.jp}
\affiliation{%
 Kansai Institute for Photon Science, National Institutes for Quantum Science and Technology (QST), 8-1-7 Umemidai, Kizugawa, Kyoto 619-0215, Japan
}%
\affiliation{%
 Photon Science Center, Graduate School of Engineering, The University of Tokyo, 7-3-1 Hongo, Bunkyo-ku, Tokyo 113-8656, Japan
}

\author{Yasushi Shinohara}
\affiliation{%
 Photon Science Center, Graduate School of Engineering, The University of Tokyo, 7-3-1 Hongo, Bunkyo-ku, Tokyo 113-8656, Japan
}
\affiliation{
NTT Basic Research Laboratories, NTT Corporation, 3-1, Morinosato-Wakamiya, Atsugi, Kanagawa 243-0198, Japan
}
\affiliation{
NTT Research Center for Theoretical Quantum Information, NTT
Corporation, 3-1 Morinosato Wakamiya, Atsugi, Kanagawa 243-0198,
Japan
}
\author{Kenichi L. Ishikawa}
\email[corresponding author:]{ishiken@n.t.u-tokyo.ac.jp}
\affiliation{%
 Photon Science Center, Graduate School of Engineering, The University of Tokyo, 7-3-1 Hongo, Bunkyo-ku, Tokyo 113-8656, Japan
}
\affiliation{%
 Department of Nuclear Engineering and Management, Graduate School of Engineering, The University of Tokyo,7-3-1 Hongo, Bunkyo-ku, Tokyo 113-8656, Japan
}%
\affiliation{%
 Research Institute for Photon Science and Laser Technology, The University of Tokyo, 7-3-1 Hongo, Bunkyo-ku, Tokyo 113-0033, Japan
}
\affiliation{%
Institute for Attosecond Laser Facility, The University of Tokyo, 7-3-1 Hongo, Bunkyo-ku, Tokyo 113-0033, Japan
}




\date{\today}

\begin{abstract}
The description of electron-electron scattering presents challenges in the microscopic modeling of the interaction of ultrashort intense laser pulses with solids. 
We extend the semiclassical approach based on the Vlasov equation [Phys. Rev. B {\bf 104}, 075157(2021)] to account for dynamic electron-electron scattering by introducing the Vlasov-Uehling-Uhlenbeck (VUU) equation.
We further couple the VUU equation with Maxwell's equations to describe the laser pulse propagation.
We apply the present approach to simulate laser-electron interactions in bulk and thin-film aluminum, focusing on energy absorption and transport.
Our calculation results reveal that electron-electron scattering affects energy absorption more significantly under p-polarization than under s-polarization, highlighting the role of the non-uniform surface potential. 
Our simulations also show that the energy transport extends beyond the optical penetration depth, which is consistent with observations in previous laser ablation experiments.
The developed Maxwell-VUU approach is expected to advance the understanding of intense laser-material interactions not only as a cost-effective alternative to the time-dependent density functional theory (TDDFT), but also by incorporating fermionic two-body collisions whose description is limited in TDDFT.

\end{abstract}

\pacs{Valid PACS appear here}
\maketitle


\section{\label{sec:introduction}INTRODUCTION}

The interaction of ultrashort intense laser pulses with solids is relevant to various research fields ranging from high-harmonic generation \cite{Shambhu,Vampa,You,Ndabashimiye,Morimoto} to material processing \cite{Zhigilei,Ilday,Audouard,Betz,Schmidt,Tiinnermann,Mourou,Wang,Gamaly}.
The ultrafast laser material processing, which suppresses heat-affected damage, starts from the energy transfer from the laser to the material by electron excitation, followed by instability of lattice system or energy transfer from the heated electrons to the lattice.
Next, the material undergoes phase and/or structural transition\cite{Medvedev}, leaving a change of the optical properties or a defect behind \cite{Mazur}, which eventually leads to irreversible damage, drilling, or structuring \cite{Mirza,Rethfeld,Rudenko,Chimier,Lorazo,Thorstensen,Kondo,Upadhyay,Ivanov,Itina2,Garrison,Sakabe,Ishino}.

The comprehensive modeling of laser material processing is highly complex, multi-scale in both time and space, multi-phase (solid, fluid, plasma, atomic gas, etc.), and possibly accompanied by chemical reactions.
Plasma or continuum models \cite{Zhigilei,Audouard,Matzen,Lehman,Wu,Silaeva} have been employed to describe and simulate the physical initial processes, advancing fundamental understanding.
However, they have difficulties in examining most initial transient dynamics before the system reaches local thermodynamic equilibrium.

It has become possible to describe the non-equilibrium electron dynamics driven by strong electromagnetic waves using the time-dependent density-functional theory (TDDFT) \cite{salmon,Yabana,Magyar,Otobe,Tani2022-yy} or time-dependent density-matrix methods \cite{DM,Hirori,Sanari}.
TDDFT is a first-principles method that strikes a good balance between accuracy and computational feasibility.
However, the computational cost is still very high for performing repeatedly to find optimal laser parameters. Another significant challenge with TDDFT is to find an exchange-correlation functional that can properly incorporate dynamical electronic correlations such as electron-electron scattering, though no systematic improvement exists.

Recently, we have developed a semiclassical method based on the Vlasov equation as a cost-effective alternative to TDDFT for studying electron dynamics in bulk solid metals under intense laser fields \cite{Tani2021-vf}.
Starting from ground state distribution obtained by solving Thomas-Fermi model, time evolution is calculated by pseudoparticle method. 
The optical spectra, including optical conductivity, dielectric function, complex refractive index, and reflectivity are evaluated by linear response calculation using impulse response, have been shown to agree well with TDDFT and experimental data.
The absorbed energy, which has been evaluated as work done by laser field, has shown a linear dependence on the incident laser intensity. This trend have agreed with TDDFT results in a certain laser fluence region.

In this paper, we further extend the Vlasov-based approach \cite{Tani2021-vf} in two ways.
First, we include dynamical electron-electron scattering, which is recognized as a vital process in metallic nano-clusters \cite{Kohn,Heraud2021-qj} and induces avalanche ionization in dielectrics and semiconductors in the context of laser material processing.
Second, we consider the propagation effect of electromagnetic waves, which is crucial due to the fact that laser pulses typically used for laser machining can only penetrate the skin of metals.

To achieve the above purposes, in this work, we introduce the Vlasov-Uehling-Uhlenbeck (VUU) equation together with Maxwell's equations to describe coupled laser-electron dynamics in a metal thin film under intense laser fields. The VUU equation is an extension of the Vlasov equation taking into account fermionic two-body collision \cite{Uehling1933-dq}, which is difficult to treat with TDDFT.
The VUU equation is numerically solved by pseudoparticle method \cite{Tani2021-vf}, which represents the electron cloud as an assembly of classical pseudoparticles whose motion is governed by Newton equation. The scattering term is evaluated by Monte Carlo sampling \cite{Kohn}.
Then, we couple the VUU equation with Maxwell's equations, numerically solved with finite differential method.
We apply the present method to a thin aluminum film.
First, our VUU simulations reveal that the effect of electron-electron scattering on absorbed energy is significant for p-polarized light while it is minor for s-polarization.
Second, our Maxwell-VUU simulations show that the energy deposition reaches deeper than would be expected by laser pulse penetration alone.

The present paper is organized as follows.
Section~\ref{sec:methods} describes our simulation methods. We review the VUU
equation and Maxwell's equations, and describe our numerical implementations with the
periodic boundary condition. In Sec.~\ref{sec:results} we describe numerical application of VUU simulation and Maxwell-VUU simulation to bulk or thin aluminium. The conclusions are given in Sec.~\ref{sec:conclusions}.

\section{\label{sec:methods}METHODS}
\subsection{\label{subsec:Vlasov equation}The Vlasov-Uehling-Uhlenbeck (VUU) equation}

The VUU equation describes the dynamics of the electron distribution $f(\mathbf{r},\mathbf{p},t)$ in the phase space as,
\begin{align}
    \frac{\partial}{\partial t}&f(\mathbf{r},\mathrm{\mathbf{p}},t) + \frac{\mathrm{\mathbf{p}}}{m}\cdot \nabla_{\mathbf{r}}f(\mathbf{r},\mathrm{\mathbf{p}},t) \notag\\
    &= \nabla_{\mathbf{r}}V_{\mathrm{eff}}[n_e(\mathbf{r},t)]\cdot \nabla_{\mathrm{\mathbf{p}}}f(\mathbf{r},\mathrm{\mathbf{p}},t) + I_{\mathrm{UU}}(\mathbf{r},\mathrm{\mathbf{p}},t), \label{Vlasov}
\end{align}
where $\mathrm{\mathbf{r}}$, $\mathrm{\mathbf{p}}$, and $m$ are the electron position, canonical momentum, mass respectively, $V_{\mathrm{eff}}$ the effective mean-field potential, and $I_{\mathrm{UU}}$ the two-body collision integral (see below) \cite{Kohn}, which was absent in our previous work \cite{Tani2021-vf}.

The effective potential $V_{\mathrm{eff}}$ is a functional of the electron density $n_e({\bf r},t)$ and decomposed into,
\begin{equation}
    V_{\mathrm{eff}}[n_e(\mathbf{r},t)] = V_{\mathrm{Coulomb}}[n_{\mathrm{e}}(\mathbf{r},t)] + V_{\mathrm{xc}}[n_e(\mathbf{r},t)] + V_{\mathrm{ext}}(\mathbf{r},t), \label{Veff}
\end{equation}
with the exchange-correlation potential $V_{\mathrm{xc}}$, external field potential $V_{\rm ext}$, and, 
\begin{equation}
    V_{\mathrm{Coulomb}}[n_e(\mathbf{r},t)] = \sum_{i}V_{\mathrm{ps}}(\mathbf{r}-\mathbf{r}_i) + V_{H}[n_e(\mathbf{r},t)],
\end{equation}
where $i$, $V_{\mathrm{ps}}$ and $V_H$ denote the label of ions and the spherically symmetric ionic pseudopotential and the electron-electron Hartree potential, respectively.
In this work, we employ the modified Heine-Abarenkov local pseudo potential for $V_{\mathrm{ps}}$~\cite{Vps,Tani2021-vf}. 
$V_H$ is evaluated by solving the Poisson equation,
\begin{equation}
    \label{eq:Poisson}
    \Delta V_H [n_e(\mathbf{r},t)] = -4\pi e n_e(\mathbf{r},t).
\end{equation}
The laser-electron interaction is described in the length gauge,
\begin{equation}
    V_{\mathrm{ext}}(\mathbf{r},t) = -e\mathbf{E}(t)\cdot \mathbf{r}, \label{Er}
\end{equation}
within the dipole approximation, where ${\bf E}$ denotes the laser electric field vector.
We use the local-density approximation (LDA) by Perdew and Zunger \cite{PZ} for the exchange-correlation potential $V_{\mathrm{xc}}$.

In our previous study \cite{Tani2021-vf} we have solved the (collisionless) Vlasov equation or Eq.~\eqref{Vlasov} with $I_{\mathrm{UU}}$ switched off, which can be derived as the leading order of a semiclassical $\hbar$ expansion of the time-dependent Kohn-Sham equation \cite{Tani2021-vf} and also be viewed as a dynamic generalization of the static Thomas-Fermi model. Although electron-electron scattering is, in principle, included in $V_{\mathrm{xc}}$, but is significantly limited in practice.
Therefore, we explicitly introduce the collision integral $I_{\mathrm{UU}}$ in Eq.~\eqref{Vlasov}.

To solve Eq.~(\ref{Vlasov}) numerically, we employ the pseudoparticle method, where the distribution $f(\mathbf{r},\mathbf{p},t)$ is treated as a set of many pseudoparticles,
\begin{equation}
    f(\mathbf{r},\mathbf{p},t) \equiv \frac{1}{N_s} \sum_{i=1}^{N_{\mathrm{pp}}} g_r(\mathbf{r}-\mathbf{r}_i(t))g_p(\mathbf{p}-\mathbf{p}_i(t)).
\end{equation}
Here $\mathbf{r}_i, \mathbf{p}_i$ are the position and canonical momentum of each pseudoparticle labeled by $i$.
The total number of pseudoparticles $N_{\mathrm{pp}}$ is expressed as $N_{\mathrm{pp}}=N_s N_e$ where $N_s, N_e$ are the number of pseudoparticles per electron and the total number of electrons contained in the simulation box $\Omega$, respectively.
$N_s$ is set to $10000$ as it is sufficient to suppress statistical noise level \cite{Tani2021-vf}.
$g_r, g_p$ denote smoothing kernel functions in real and momentum space, respectively, of Gaussian forms,
\begin{align}
    g_r(\mathbf{r})&=\sum_{\{ \mathbf{G} \}} \frac{1}{\pi^{3/2}d_r^3} \exp(-|\mathbf{r}+\mathbf{G}|^2/d_r^2), \\
    g_p(\mathbf{p})&= \frac{1}{\pi^{3/2}d_p^3} \exp(-|\mathbf{p}|^2/d_p^2),
\end{align}
where $d_r, d_p$ are smoothing widths in real and momentum space, respectively.
Within the pseudoparticle method, the collision integral $I_{\mathrm{UU}}$ is treated as stochastic collision events among pseudoparticles taking place when accepted by Monte Carlo sampling \cite{Kohn}.
At each time step, we randomly pick up a pair of pseudoparticles, whose momenta are $(\mathbf{p}_1,\mathbf{p}_2)$. If their distance is smaller than “impact parameter" $b=\sqrt{(\sigma_{\mathrm{tot}}/\pi N_s)}$, we randomly choose a new momentum state after elastic collision $(\mathbf{p}_1,\mathbf{p}_2)\to(\mathbf{p}'_1,\mathbf{p}'_2)$, which is accepted by the rate $(1-f_{\mathbf{p}'_1})(1-f_{\mathbf{p}'_2})$ that accounts for Pauli blocking, where $f_{\mathbf{p}}$ denotes the local momentum occupation. Here $\sigma_{\mathrm{tot}}$ denotes total scattering cross section that is derived as that by screened Coulomb potential which solely depends on the electron density (details are given in Ref.~\cite{Kohn}).
If the collision event mentioned above is not accepted, a new pair will be sampled, and the same process will be repeated. However, once a pseudoparticle is accepted in a scattering event, it will not be used for another scattering event within the same time step.

We employ a real-space simulation box $\Omega$ for the electron system, on which the periodic boundary condition is imposed \cite{Tani2021-vf}.
Numerical implementation for the ground state calculation, described by the Thomas-Fermi model, and the time propagation for the left-hand side in Eq.~(\ref{Vlasov}) is described in Ref.~\cite{Tani2021-vf} in details.
The charge current density $\mathbf{J}(t)$ averaged over the simulation box $\Omega$ is given by,
\begin{equation}
    \mathbf{J}(t) = \frac{1}{|\Omega|}\iint_{\Omega} \left(-e \frac{\mathbf{p}}{m}\right) f(\mathbf{r},\mathbf{p},t) \dd \mathbf{r} \dd \mathbf{p},
    \label{eq:current density}
\end{equation}
where $|\Omega|$ denotes the volume of the simulation box. The energy current is defined in an analogous way,
\begin{equation}
    \mathbf{Q}(t) = \frac{1}{|\Omega|}\iint_{\Omega} \left(\frac{\mathbf{p}^2}{2m} \frac{\mathbf{p}}{m}\right) f(\mathbf{r},\mathbf{p},t) \dd \mathbf{r} \dd \mathbf{p}.
    \label{eq:current density}
\end{equation}

\subsection{\label{subsec:maxwell}Maxwell's equations}

\begin{figure}[tb]
  \centering
  \includegraphics[keepaspectratio,width=\hsize]{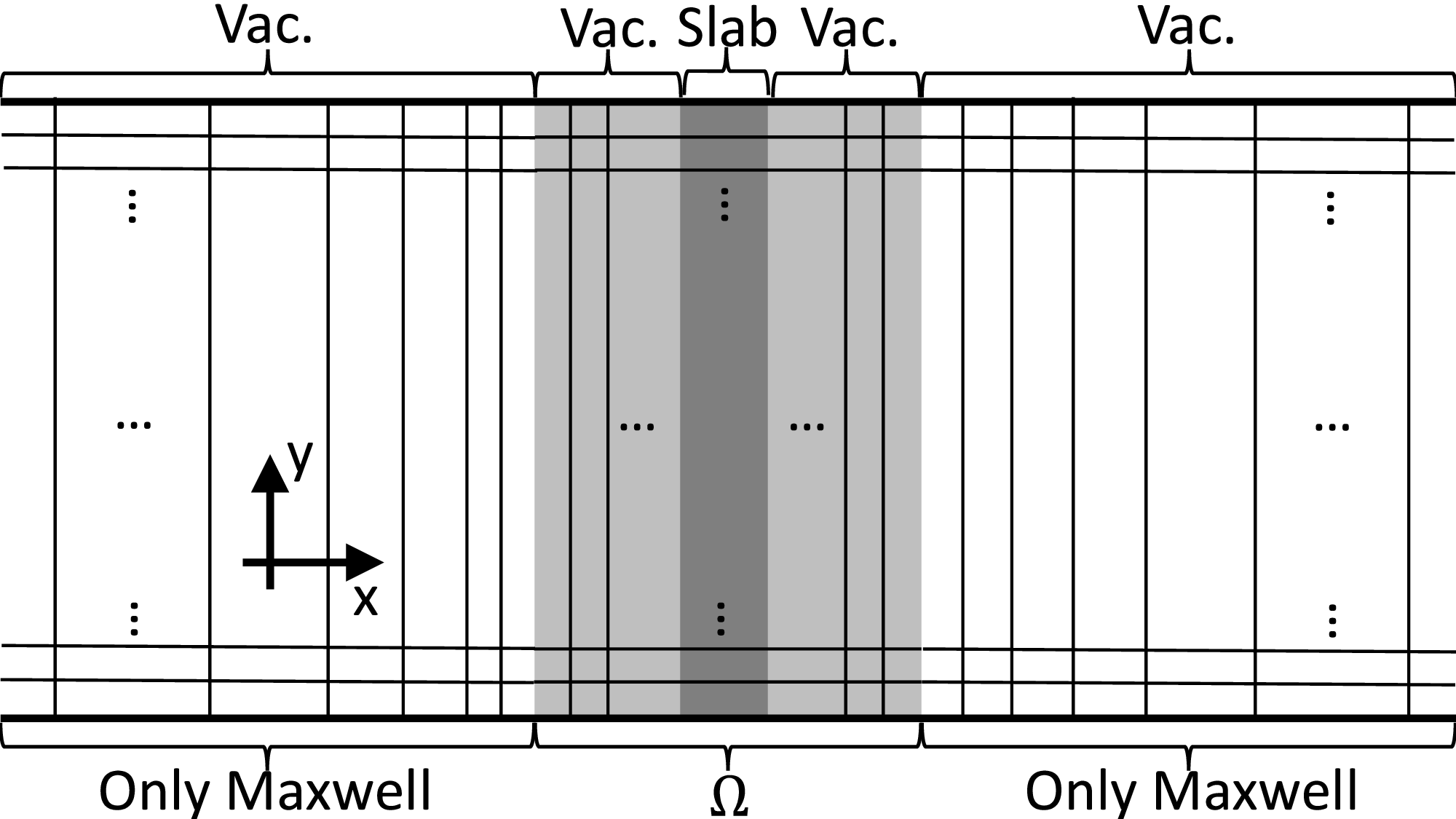}
  \caption[overview] {Model geometry. $\Omega$, which is composed of the slab region and the vacuum region, is the simulation box for both the electronic system and the Maxwell system. The Vacuum domain outside $\Omega$ is only for the Maxwell system.}
  \label{overview}
\end{figure}

We consider normal incidence of a linearly polarized laser pulse.
The Maxwell equation to be solved is,
\begin{equation}
    \left( \frac{1}{c^2}\frac{\partial^2}{\partial t^2} - \Delta \right)\mathbf{A}(\mathbf{r},t)+\frac{\partial}{\partial t} \nabla V_{\mathrm{H}}(\mathbf{r},t) = \mu_0\mathbf{J}(\mathbf{r},t), \label{1dMaxwell}
\end{equation}
where $c$ denotes the light speed, $\mathbf{A}(\mathbf{r},t)=-\int_{\infty}^{t} \mathbf{E}(\mathbf{r},\tau)d\tau$ the vector potential, $\mathbf{E}(\mathbf{r},t)$ the electric field, and $\mathbf{J}(\mathbf{r},t)=\int\left(-e \frac{\mathbf{p}}{m}\right) f(\mathbf{r},\mathbf{p},t) \dd \mathbf{p}$ the local current density at each grid point.
Note that field quantities related to Eq.~(\ref{1dMaxwell}), $\mathbf{A},\mathbf{E},\mathbf{J}$, are of the Maxwell system.
The VUU equation (\ref{Vlasov}) and Maxwell equation (\ref{1dMaxwell}) are coupled through the electric field and the current density.
The electric field $\mathbf{E}(\mathbf{r},t)=-\dot{\mathbf{A}}(\mathbf{r},t)$, obtained from Eq.~(\ref{1dMaxwell}), is used as an input to Eq.~(\ref{Vlasov}).
Pseudoparticles feel the electric field at the nearest grid point.
The initial wave form in vacuum is given as a spatial distribution of the vector potential. At the boundary of the simulation box for the Maxwell equation, which is extended longer than $\Omega$ along the optical axis, Mur's absorbing boundary conditions are imposed to suppress unphysical reflection \cite{Mur1981-ca}.
Figure~\ref{overview} shows a schematic diagram in x-y plane.
While the VUU and Maxwell equations share the grid points inside $\Omega$, the latter is solved also in vacuum layers attached to both sides of $\Omega$.

\section{\label{sec:results} RESULTS AND DISCUSSIONS}
\subsection{\label{subsec:impulse response}Linear response}
\begin{figure}[tb]
  \centering
  \includegraphics[keepaspectratio,width=\hsize]{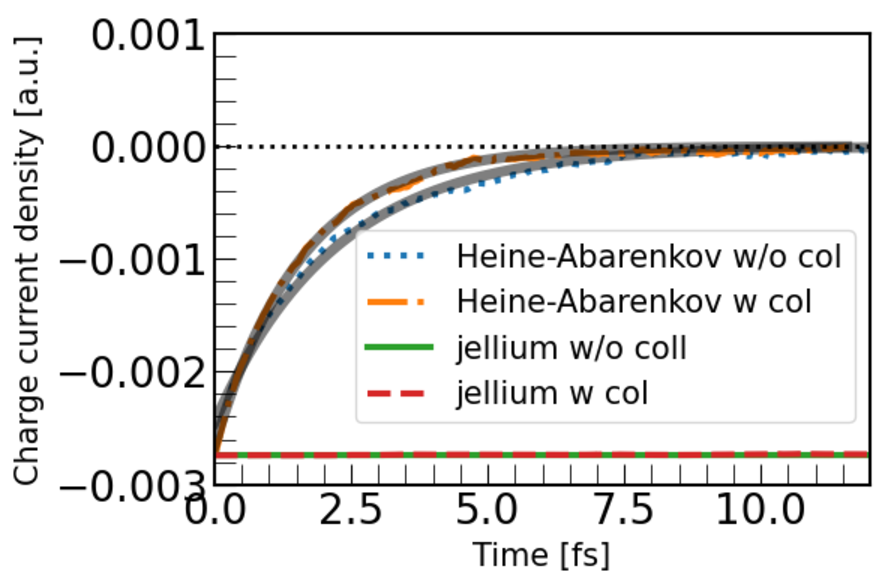}
  \caption[jellium]{Time evolution of charge current density after impulsively added momentum $\Delta \mathbf{p}$ at $t=0$. Blue dotted and orange dash-dotted lines indicate Vlasov and Vlasov-UU simulation results with the modified Heine-Abarenkov potential. Green solid and red dashed lines represent Vlasov and Vlasov-UU simulation results with the jellium model. The black translucent lines represent the fitting results using Eq.~(\ref{J_tau}) of the simulation curve with the Heine-Abarenkov type potential.}
  \label{jellium}
\end{figure}

Let us first examine the linear optical response by applying an impulse electric field which is equivalent to setting initial pseudoparticle momenta $\mathbf{p}_i = \mathbf{p}_i^{\mathrm{GS}} + \Delta \mathbf{p}$ shifted from the ground-state values $\mathbf{p}_i^{\mathrm{GS}}$ by a small amount $\Delta \mathbf{p} = (0, 0, 0.1) \ \mathrm{a.u.}$ \cite{Tani2021-vf}.
In this simulation, we solely use the simulation box $\Omega$ on which the periodic boundary condition is applied for all the three directions with no vacuum region.
Figure~\ref{jellium} shows the comparison of the calculated current density among the Vlasov and Vlasov-UU simulations with the modified Heine-Abarenkov potential and a jellium model, where a uniform positive charge is used instead of ionic pseudopotentials.

The time evolution of the current density is fitted by,
\begin{equation}
    J_z(t)=J_z^0\exp(-t/\tau), \label{J_tau}
\end{equation}
where fitting parameters are the initial value $J_z^0$ and damping constant $\tau$.
$\tau$ is 2.2 and 1.6 fs for the collisionless (Vlasov) and the collisional (Vlasov-UU) case, respectively. While both collisionless and collisional dynamics involve damping, the difference in $\tau$ implies that our simulations include momentum relaxation due to electron-electron scattering. It should be noted that the electron scattering from the ionic potential, which leads to the damping as is seen in Fig.~\ref{jellium}, is a classical analog of Umklapp scattering \cite{Abrikosov}. In fact, in the calculation using the jellium model, we observe current damping for neither collisionless nor collisional calculations.

\subsection{\label{subsec:energy absorption} Energy transfer from laser to electrons}
Let us next investigate the energy absorption of an aluminum film from the laser pulse. The electric field is assumed to be spatially uniform and given by,
\begin{equation}
    E(t)=E_0\sin (\omega t) \cos^2\left[\frac{t\pi}{T}\right] \ \left(-\frac{T}{2}\le t\le \frac{T}{2}\right), \label{e_field}
\end{equation}
where $E_0$ denotes the field amplitude, $\hbar \omega$ the photon energy, and $T$ the (foot-to-foot) full pulse duration. The laser intensity profile has a full width at half maximum (FWHM) duration of approximately $0.36T$. 
In this subsection, Eq.~(\ref{Vlasov}) is solved without coupling with Eq.~(\ref{1dMaxwell}).
We solely use simulation box $\Omega$ on which the periodic boundary condition is applied for the electron system. $\Omega$ contains 4 nm vacuum region on both sides of the film along the optical axis.

Here we consider a central wavelength of 400 nm, peak intensity of $1 \mathrm{TW/cm}^2$, and pulse width of 32.4 fs in FWHM (corresponding to 90 fs in foot-to-foot pulse width). This laser pulse is applied to a 4nm-thick aluminum slab in a vacuum.

\begin{figure}[tb]
  \centering
  \includegraphics[keepaspectratio,width=\hsize]{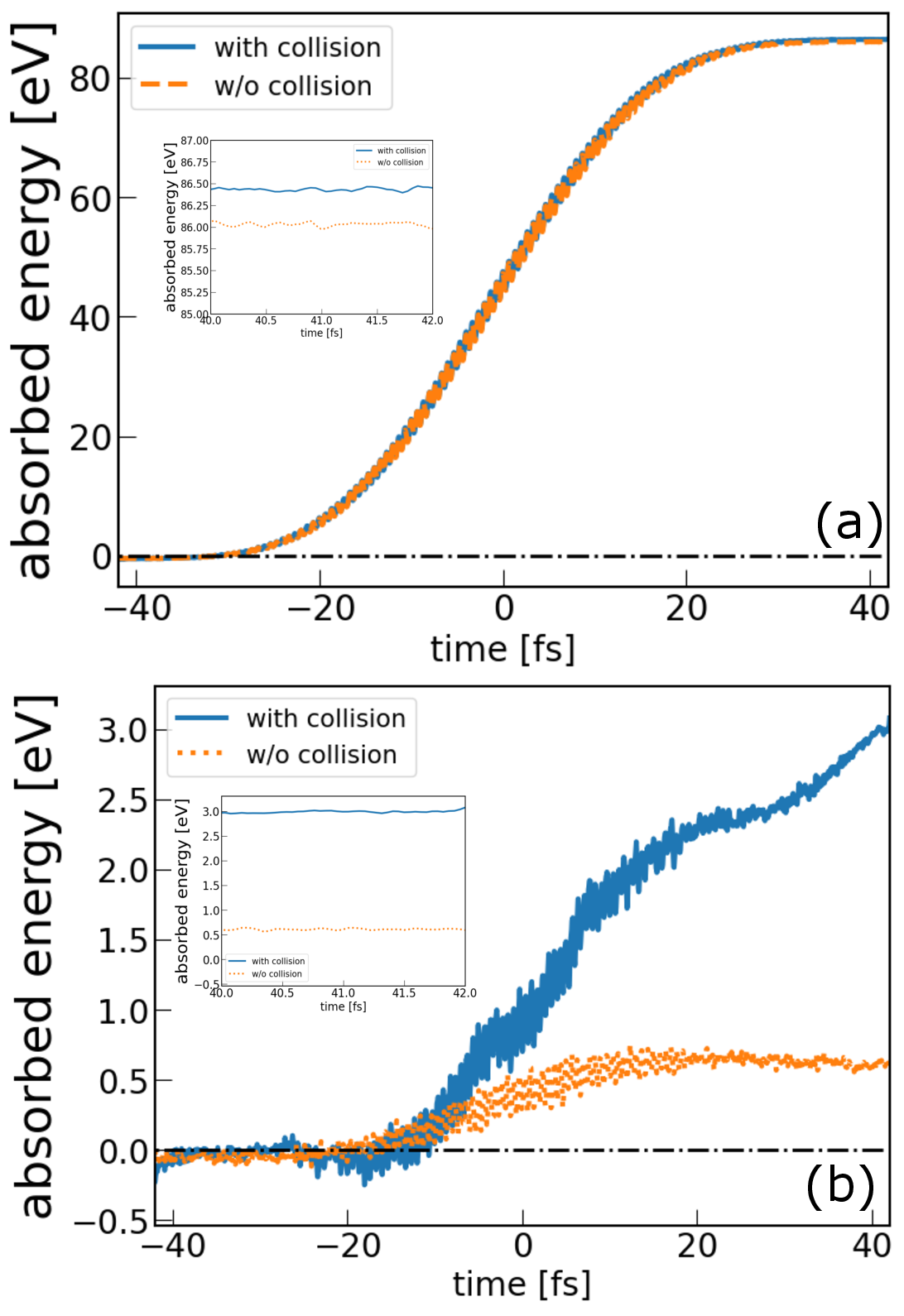}
  \caption[sp-comparison]{Absorbed energy under (a) s-polarized light and (b) p-polarized light. Blue solid and orange dashed lines are results with and without electron-electron scattering. Insets are magnified view of the time period around the end of the laser pulse.}
  \label{sp-comparison}
\end{figure}

We evaluate the absorbed energy as the difference in totoal electron energy from the initial state. Figure~\ref{sp-comparison} shows the time evolution of the absorbed energy under s- and p-polarized laser pulses.
Increase in absorbed energy due to electron-electron scattering is larger for p-polarization ($\sim 2.5$ eV) than for s-polarization ($\sim 0.5$ eV). This observation can be attributed to two factors: the scattering rate and the contribution from potential scattering.
In the p-polarized case, the overall electron motion is perpendicular to the surface. The difference in the density of states between the surface and the bulk may lead to a temporally low occupancy, resulting in frequent electron-electron scattering due to the lower Pauli blocking rate.
For the simulations with electron-electron scattering [blue solid line in Fig.~\ref{sp-comparison}(b)], the absorbed energy still increases during the trailing edge of the pulse, while it is constant after the pulse [inset of Fig.~\ref{sp-comparison}(b)].
A possible reason is that once the electrons are excited, the occupation in the momentum space becomes far from the step function, then electron-electron scattering occurs more easily after thermalization, leading to an increase in the absorption rate even under a weak laser field.
A common process for both s-polarization and p-polarization is scattering from the ionic core potentials.
However, the surface electrons are scattered by the surface Coulomb potential, further promoting collisional dissipation under p-polarized laser fields.
It is easy to show that elastic collision of two particles at $t=t_{\mathrm{col}}$ under a uniform external potential, upon which the momentum chenge is described by $(\mathbf{p}_1(t_{\mathrm{col}}),\mathbf{p}_2(t_{\mathrm{col}})) \to (\mathbf{p}_1'(t_{\mathrm{col}}),\mathbf{p}_2'(t_{\mathrm{col}}))$, does not change the final total energy,
\begin{align}
    \frac{1}{2m}(\mathbf{p}_1&+\mathbf{A}_{\mathrm{col}})^2 + \frac{1}{2m}(\mathbf{p}_2+\mathbf{A}_{\mathrm{col}})^2 \notag \\
    &= \frac{1}{2m}(\mathbf{p}_1'+\mathbf{A}_{\mathrm{col}})^2 + \frac{1}{2m}(\mathbf{p}_2'+\mathbf{A}_{\mathrm{col}})^2,
    \label{energy-conservation}
\end{align}
where $\mathbf{A}_{\mathrm{col}}$ denotes the laser vector potential at the time of the collision.
On the other hand, the acute non-uniformity of the surface Coulomb potential experienced by the returning electrons under p-polarization leads to a larger increase in the final total energy.

\subsection{\label{subsec:energy transport} Energy transport}
In this subsection, we utilize the Maxwell-Vlasov(-UU) simulation to investigate how the electrons are excited in an aluminum film which has 16 nm thickness under a propagating linearly polarized laser pulse whose initial electric field profile is given by,
\begin{equation}
    E(x,t=0)=E_0 \sin{\left[ \frac{\omega}{c} x\right]}\sin^2\left(\frac{x}{cT}\pi\right) \ (0\le x\le cT).
\end{equation}
We use simulation box $\Omega$ on which periodic boundary condition is applied for electron system. $\Omega$ contains 4nm vacuum regions on both sides of the film along the optical axis. The simulation box for the Maxwell equation has additional vacuum layers of $4\mu$m thickness attached to both sides of $\Omega$.

\begin{figure}[tb]
  \centering
  \includegraphics[keepaspectratio,width=0.95\hsize]{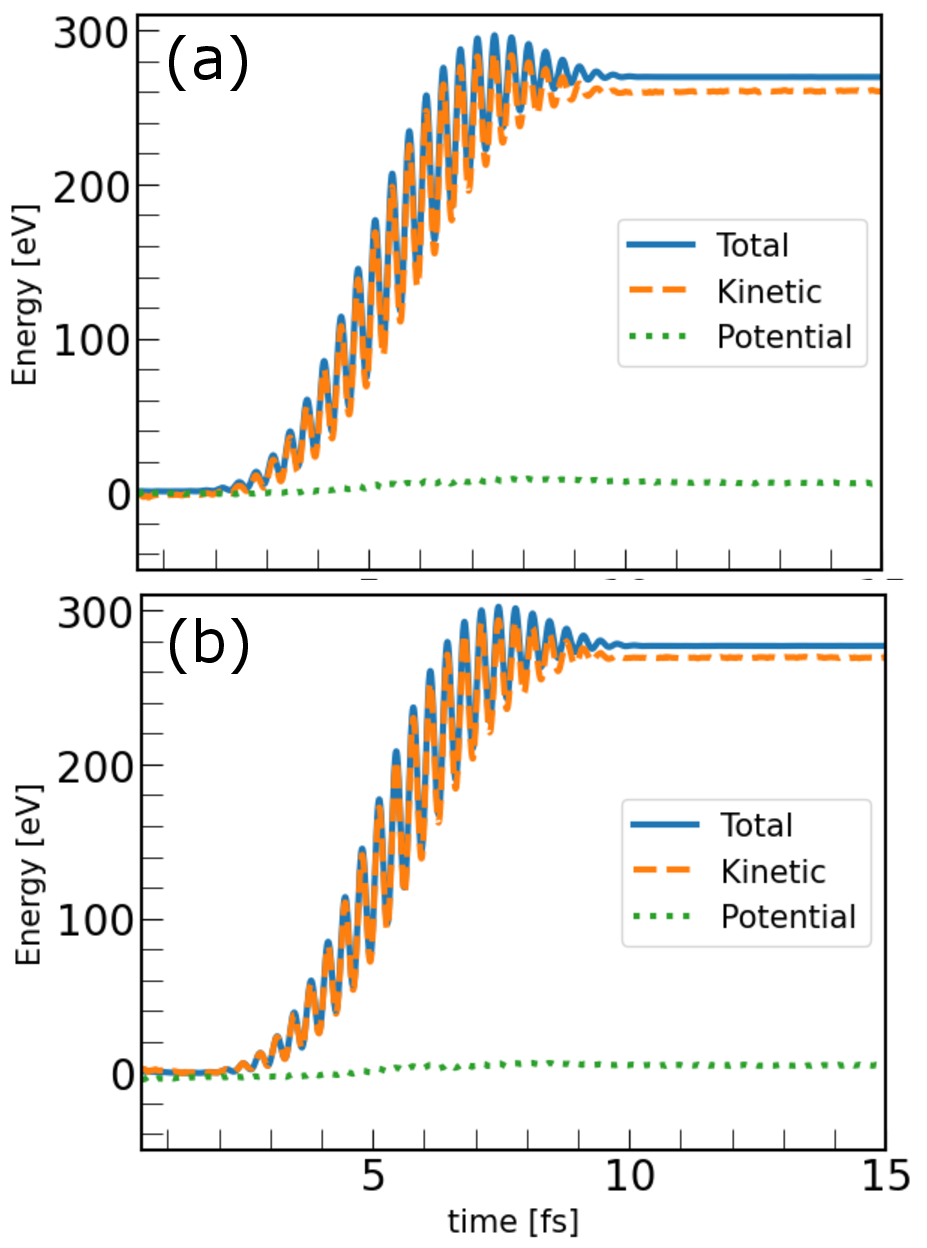}
  \caption[totkinpot]{Time evolution of energy gain is illustrated. Total energy is ploted in solid blue line, kinetic energy in dashed orange line, and potential energy in dotted green line. (a) without collision, (b) with collision.}
  \label{tot_kin_pot}
\end{figure}

\begin{figure}[tb]
  \centering
  \includegraphics[keepaspectratio,width=0.95\hsize]{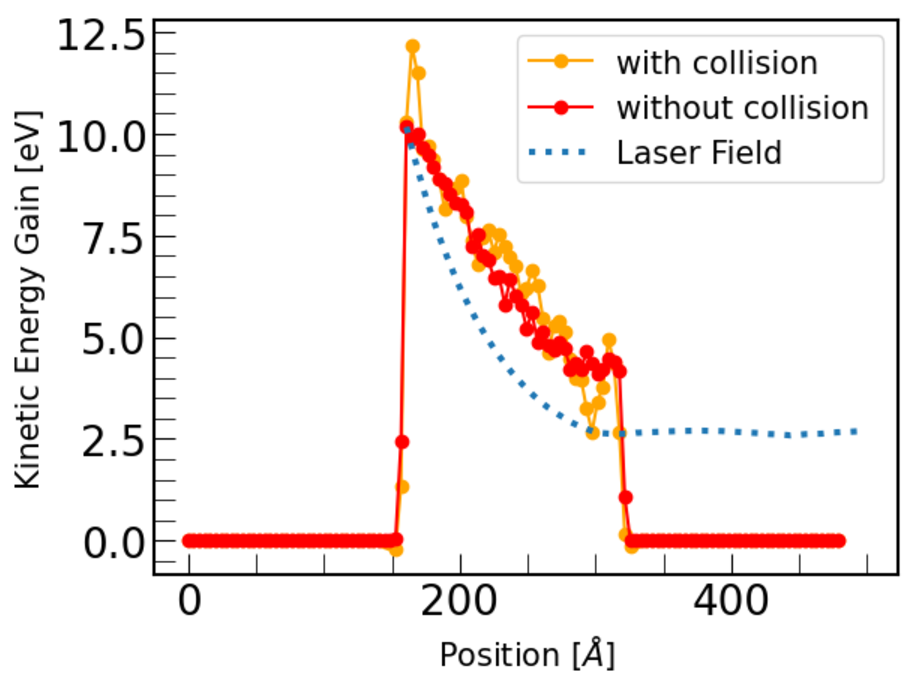}
  \caption[kindistribution]{Kinetic energy gain distribution calculated by collisionless simulation in red line and by collisional simulation in orange, and scaled maximum laser field intensity along optical axis in blue dashed line.}
  \label{kin_distribution}
\end{figure}

We consider a laser pulse with a central wavelength of 200 nm, peak intensity of $100\, \mathrm{TW/cm}^2$, and FWHM pulse width of 3.6 fs (10 fs foot-to-foot), normally incident to a 16nm-thick aluminum slab in a vacuum.
Figure~\ref{tot_kin_pot} shows the temporal evolution of the total, kinetic, and potential energies for collisional and collisionless simulations.
The total energy is the sum of the kinetic and the potential energies. 
We can see that the kinetic energy gain is dominant for both collisional and collisionless cases, and, thus, let us focus on its dynamics. 
Figure~\ref{kin_distribution} presents the depth distribution of the kinetic energy just after irradiation. The normalized laser peak intensity at each position along optical axis is also plotted.
The kinetic energy distribution is not proportional to the depth profile of the laser intensity but penetrates deeper in the sample, which means energy transport along the optical axis is taking place during short time of laser pulse irradiation.
The difference between the optical and electron heat penetration depths has been reported experimentally by \citet{Miyasaka2015-pw}.
Our results support such observations based on a microscopic theoretical framework.

\begin{figure}[tb]
  \centering
  \includegraphics[keepaspectratio,width=0.95\hsize]{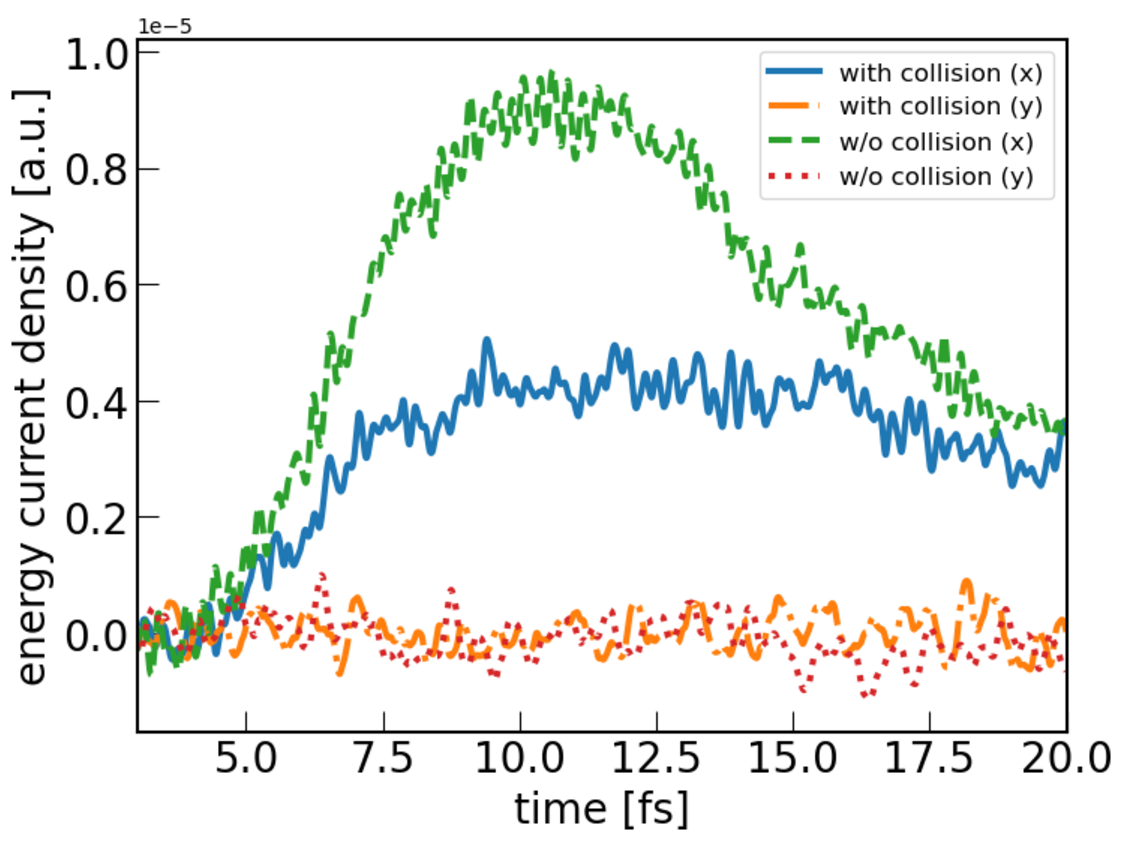}
  \caption[energycurrent]{Time evolution of energy current density in x- and y-direction. Blue solid and green dashed lines indicate x-component with and without electron-electron scattering, orange dash-dotted and red dotted lines does y-component with and without electron-electron scattering.}
  \label{energy_current}
\end{figure}

Let us next quantitatively explore the energy flow given by Eq.~(\ref{eq:current density}) and how electron-electron scattering affects its behavior.
Figure~\ref{energy_current} shows the temporal evolution of the energy current density averaged on the whole space including the vacuum domains.
The energy flow exists only along the optical axis ($x$ direction), with negligible net lateral components.
In addition, we see that the energy current along the optical axis is reduced by electron-electron scattering, which partially redirects the electron motion to transverse directions.
Despite this observation, the distribution of the kinetic energy gain in Fig.~\ref{kin_distribution} is surprisingly similar for the collisional and collisionless cases.
Indeed, the ``center of mass (c.m.)" of the distribution with collision (225.1\AA) is only slightly deeper than that without collision (223.3\AA).
On the other hand, c.m.~of the distribution of the total energy gain gets deeper from 298.5\AA~to 318.7\AA, which may suggest that the kinetic energy is eventually transformed to the potential energy.

\section{\label{sec:conclusions} Conclusions}

We have extended the semiclassical formalism based on the Vlasov equation \cite{Tani2021-vf} to incorporate electron-electron scattering as well as propagating laser field dynamics.
The electronic distribution function is expressed by a set of pseudoparticles.
Electron-electron scattering is treated as collision events between pseudoparticles, where the Pauli blocking factor is taken into account in the acceptance rate.
The propagation of the laser field is calculated by solving the Maxwell equations.

We have applied the developed approach to crystalline and slab aluminum.
After the application of an impulse electric field to bulk aluminum, the Vlasov-UU calculation leads to a faster damping of the charge current than the collisionless Vlasov calculation, due to the momentum relaxation by interelectronic scattering.
The investigation of the effect of the laser polarization on energy absorption in an aluminum film has shown that electron-electron scattering increases the absorbed energy, especially under p-polarization, rather than under s-polarization. This result indicates an important role of the non-uniform surface Coulomb potential.
Furthermore, our Maxwell-Vlasov(-UU) simulations have demonstrated that energy is transported deeper into the aluminum film than would be expected from the laser penetration depth alone. While electron-electron scattering suppresses the energy flow along the optical axis, it does not significantly alter the spatial distribution of the kinetic energy gain, implying the transformation of the kinetic energy into the potential energy.

These extensions allow the investigation of electron dynamics in bulk and slab metals under ultrashort intense laser pulses while incorporating the effects of electron-electron scattering as well as laser pulse propagation. The former extension offers an advantage over TDDFT, since the description of electron-electron collisions in TDDFT is limited. The latter extension overcomes the limitation of our previous Vlasov simulator, since laser propagation is potentially important when the object size is not sufficiently smaller than the laser wavelength. The next step will be to incorporate molecular dynamics, which plays a crucial role after strong electron excitation, leading to laser-induced irreversible damage.

\section*{\label{sec:acknowledgements} ACKNOWLEDGEMENTS}
We wish to express our gratitude to Kazuhiro Yabana for fruitful discussions. 
This research was supported by MEXT Quantum Leap Flagship Program (MEXT Q-LEAP) Grant Number JPMXS0118067246. 
This research was also partially supported by JSPS KAKENHI Grant Number JP20H05670, JP22K18982, JP24K01224, and JP24H00427, and JST CREST under Grant Number JPMJCR16N5. 
M.T. gratefully acknowledges support from the Graduate School of Engineering, The University of Tokyo, Graduate Student Special Incentives Program. M.T. also gratefully thanks support from crowd funding platform  \it academist. \rm The numerical calculations are partially performed on supercomputers Oakbridge-CX, sekirei, ohtaka (the University of Tokyo), and SGI ICE X at Japan Atomic Energy Agency(JAEA). This research is partially supported by Initiative on Promotion of Supercomputing for Young or Women Researchers, Information Technology Center, The University of Tokyo.
\appendix
\def\thesection{\Alph{section}}
\section{\label{sec:appendixA}Finite width pseudo particles}
One should carefully derive representations of moments using one-body distribution approximated by the pseudo-particle method. Assuming a $2D$-dimensional classical phase space, 1st-order moment $A(t)$ and 2nd-order moment $B(t)$ are defined as,
\begin{align}
    A(t) &= \int_{\Gamma} \bm{\gamma} f_D(\mathbf{r},\mathbf{p},t) \dd \mathbf{r} \dd \mathbf{p}, \\
    B(t) &= \int_{\Gamma} \bm{\gamma}^2 f_D(\mathbf{r},\mathbf{p},t) \dd \mathbf{r} \dd \mathbf{p},
\end{align}
where $\bm{\gamma}=\mathbf{r},\mathbf{p}$, $f_D$ stands for a $N_p$-physical-particle one-body distribution,
\begin{equation}
    \int_{\Gamma} f_D(\mathbf{r},\mathbf{p},t) \dd \mathbf{r} \dd \mathbf{p} = N_p.
\end{equation}
Under the pseudo-particle ansatz, one-body distribution $f_D$ is represented as a function of Gaussian kernels,
\begin{align}
    f_D(\mathbf{r},\mathbf{p},t) = \frac{1}{N_s}&\sum_{i=1}^{N_{\mathrm{pp}}} \frac{1}{(\sqrt{\pi}d_{\gamma})^{2D}}\notag \\
    &\times \prod_{\gamma=r,p} \exp \left[-\left(\frac{\bm{\gamma}-\bm{\gamma}_i(t)}{d_{\gamma}}\right)^2\right], \label{fPP}
\end{align}
where $N_s=N_{\mathrm{pp}}/N_p$ denotes the number of pseudo particles per physical particle. Using the representation (Eq.~(\ref{fPP})), $A(t),B(t)$ are written as,
\begin{align}
    A(t) &= \frac{1}{N_s}\sum_{i=1}^{N_{\mathrm{pp}}} A_i(t), \\
    A_i(t) &= \int_{\Gamma} \bm{\gamma} \frac{1}{(\sqrt{\pi}d_{\gamma})^{D}} \exp \left[-\left(\frac{\bm{\gamma}-\bm{\gamma}_i(t)}{d_{\gamma}}\right)^2\right] \dd \bm{\gamma}, \\
    B(t) &= \frac{1}{N_s}\sum_{i=1}^{N_{\mathrm{pp}}} B_i(t), \\
    B_i(t) &= \int_{\Gamma} \bm{\gamma}^2 \frac{1}{(\sqrt{\pi}d_{\gamma})^{D}} \exp \left[-\left(\frac{\bm{\gamma}-\bm{\gamma}_i(t)}{d_{\gamma}}\right)^2\right] \dd \bm{\gamma}
\end{align}
$A_i(t),B_i(t)$ are analytically evaluated as,
\begin{align}
    A_i(t)&=\bm{\gamma}_i(t), \\
    B_i(t)&=\bm{\gamma}_i(t)^2+\frac{Dd_{\gamma}^2}{2},
\end{align}
therefor,
\begin{align}
    A(t) &= \frac{1}{N_s}\sum_{i=1}^{N_{\mathrm{pp}}} \bm{\gamma}_i(t), \\
    B(t) &= \frac{1}{N_s}\sum_{i=1}^{N_{\mathrm{pp}}} \left[ \bm{\gamma}_i(t)^2+\frac{Dd_{\gamma}^2}{2} \right], \\
    &=\frac{1}{N_s}\sum_{i=1}^{N_{\mathrm{pp}}}  \bm{\gamma}_i(t)^2 +  \frac{Dd_{\gamma}^2}{2}N_p.\label{Bt}
\end{align}
When one employs delta function as kernel function on the $\bm{\gamma}$ space, or $d_{\gamma}=0$, 2nd-order moment $B^{\delta}(t)$ is defined as,
\begin{align}
    B^{\delta}(t) &= \frac{1}{N_s}\sum_{i=1}^{N_{\mathrm{pp}}} B_i^{\delta}(t), \label{Bd1}\\
    B_i^{\delta}(t) &= \int_{\Gamma} \bm{\gamma}^2 \delta(\bm{\gamma}-\bm{\gamma}_i(t)) \dd \bm{\gamma}, \\
    &= \bm{\gamma}_i(t)^2. \label{Bd3}
\end{align}
Inserting Eq.~(\ref{Bd1})-Eq.~(\ref{Bd3}) into Eq.~(\ref{Bt}), one obtaines,
\begin{equation}
    B(t) = B^{\delta}(t) + \frac{Dd_{\gamma}^2}{2}N_p. \label{Bt2}
\end{equation}
On a $N_{\mathrm{pp}}$-pseudo-particles simulation, statistical mean square error of the first term in Eq.~(\ref{Bt2}) would be order of $\mathcal{O}(1) \ (N_s\to \infty)$. Although the second term is not necessarily negligible if one set a finite $d_{\gamma}$, it is just a constant shift of energy. Even if $d_{\gamma}$ is not zero, finite $d_p$ correction brings essentially no change to our results from physical context.

Actually, using a finite $d_p$, one can evaluate kinetic energy $K(t)$ of the $N_e$ electrons whose mass is $m$ as,
\begin{align}
    K(t) &= \int_{\Gamma} \frac{\mathbf{p}^2}{2m} f_3(\mathbf{r},\mathbf{p},t) \dd \mathbf{r} \dd \mathbf{p} \\
    &= \frac{1}{2m} \left[ \frac{1}{N_s}\sum_{i=1}^{N_{\mathrm{pp}}}  \mathbf{p}_i(t)^2 +  \frac{3d_{p}^2}{2}N_e \right],
\end{align}
in a 6-dimensional phase space, $D=3$. 

However, the second term in Eq.~(\ref{Bt2}) could not be missed when the absolute value of the electron energy plays a important role. For instance, heat current $J_h(t)$ is defined as,
\begin{align}
    J_h(t)&=\int_{\Gamma} \frac{\mathbf{p}^2}{2m}\frac{\mathbf{p}}{m} f_3(\mathbf{r},\mathbf{p},t) \dd \mathbf{r} \dd \mathbf{p}, \\
    &= \frac{1}{N_s}\sum_i\frac{\mathbf{p}_i(t)^2+\frac{5}{2}d_p^2}{2m}\frac{\mathbf{p}_i(t)}{m}.
\end{align}
One would see a non-negligible role of the finite $d_p$ correction, unless $d_p \ll p$ for a typical absolute value of the kinetic momentum $p$.


%


\end{document}